\documentclass[a4paper,11pt]{article}
\usepackage{pos}
\usepackage{graphicx}
\usepackage[numbers]{natbib} 
\usepackage{blindtext}
\title{Testing Gamma/Hadron Separation for Ultra-High-Energy Cherenkov Astronomy}
\ShortTitle{$\gamma$/Hadron Separation for UHE Cherenkov Astronomy}

\author*[a]{Sruthiranjani Ravikularaman}
\author[]{on behalf of the PANOSETI Collaboration}
\author[b]{Felix Riehn}
\author[b]{Dominik Elsaesser}

\affiliation[a]{Astronomisches Institut (AIRUB), Ruhr-Universität Bochum, Germany}
\affiliation[b]{Technische Universität Dortmund, Dortmund, Germany}

\emailAdd{sruthi@astro.ruhr-uni-bochum.de}

\abstract{Dark100 is a planned array of six telescopes, using the Panoramic Search for Extraterrestrial Intelligence (PANOSETI) telescope system. It will operate as an imaging atmospheric Cherenkov telescope array, with a telescope design and array layout optimized for accessing gamma rays with tens of TeV to PeV energies. The science goals of Dark100 include the search for ultra-heavy dark matter, observations of Galactic Pevatrons, and the search for ultra-fast optical transients. Rejection of background cosmic rays is key to the sensitivity of the array. We present a first study of gamma/hadron separation based on simulated gamma rays and protons, focusing on the impact of the hadronic background models used in CORSIKA.}

\ConferenceLogo{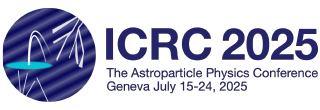}

\FullConference{39th International Cosmic Ray Conference (ICRC2025)\\
 15–24 July 2025\\
Geneva, Switzerland\\}

\begin{document}
\maketitle

\section{Introduction}

The nature of dark matter remains one of the biggest mysteries in astrophysics. A variety of dark matter candidates have been proposed, spanning different particle masses. So far, none of these candidates have been detected. Some have been ruled out by experiments \cite{magic16}, but a significant part of the mass range, between 100 TeV and tens of PeV, remains largely unexplored \cite{car23}. These particles, referred to as ultra-heavy dark matter, have become an attractive alternative as slight modifications to the dark matter paradigm unlock this range of masses \cite{tak22}. The Dark100 experiment aims to detect, or at least rule out, dark matter in this regime, with velocity-weighted cross-sections and masses that cannot be probed by existing instruments. 

\begin{figure}[htbp]
  \centering
  \includegraphics[width=0.33\linewidth]{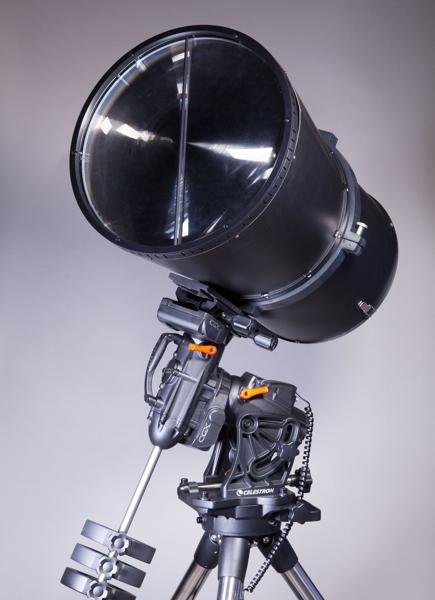}
  \includegraphics[trim=60pt 0pt 50pt 30pt, clip, width=0.52\linewidth]{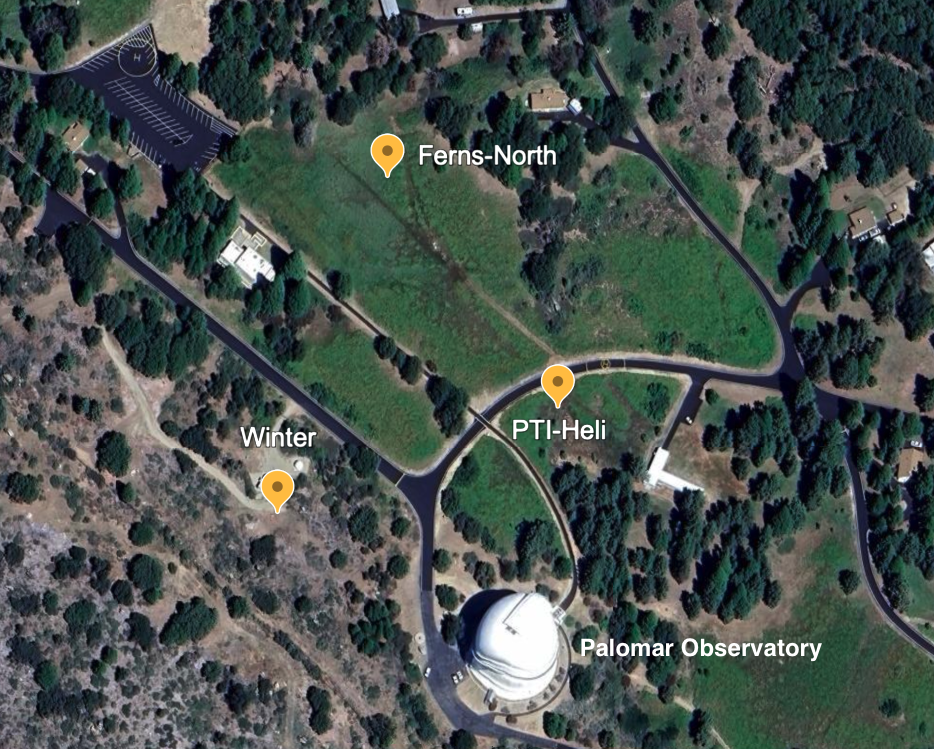}
  \caption{\textbf{Left:} Photograph of an individual PANOSETI 0.5-m Fresnel telescope. \textbf{Right:} Planned locations of the first three telescopes at Palomar.}
  \label{fig:telpal}
\end{figure}

The Dark100 experiment uses an array of Panoramic Search for Extraterrestrial Intelligence (PANOSETI) telescopes \cite{wri18}. PANOSETI telescopes (see Fig.~\ref{fig:telpal}, left) are designed to search for signs of interstellar communication or technosignatures in the optical wavelengths that could be produced by advanced extraterrestrial civilisations. Since PANOSETI telescopes are sensitive to signals on nanosecond timescales, they are well-adapted for gamma-ray astronomy above 10 TeV. Each telescope consists of four quadrant boards, each made up of a 8$\times$8 array of silicon photomultiplier (SiPM) square pixels. The camera has a total of 32$\times$32 pixels, with light focused onto it by a 0.5-m acrylic Fresnel lens. Each SiPM pixel covers $0.3^\circ$, allowing the camera to observe a $10^\circ\times 10^\circ$ section of the sky. The array, located at Palomar (see Fig.~\ref{fig:telpal}, right), will initially consist of three telescopes, with plans for future expansion.

Cherenkov light detection using PANOSETI telescopes follows the same principle used by Imaging Atmospheric Cherenkov Telescopes (IACTs). When a high-energy $\gamma$-ray or cosmic ray enters the Earth’s atmosphere, it interacts with the ambient matter and produces a shower of secondary particles.  If these secondary particles travel faster than the speed of light in the atmosphere, they emit Cherenkov light. The PANOSETI telescopes are capable of detecting these optical flashes.

The pixel size and the field of view of PANOSETI telescopes are on the same order of magnitude as those of CTA's Small Sized Telescopes. The effective collection area of a three-telescope array is roughly 0.1 km$^2$ at 10 TeV, which is approximately the effective area of VERITAS at this energy. The effective collection area is expected to increase as more telescopes are added. Previously, two PANOSETI telescopes were set up at the VERITAS site for simultaneous observations of the Crab Nebula. This experiment produced promising results, demonstrating the capability of PANOSETI telescopes to observe ultra-high-energy $\gamma$-rays \cite{mai22, kor23}. In addition, they are cost-effective and relatively quick and easy to install. With the deployment of the first telescopes approaching, we present the results of a preliminary study on the effectiveness of $\gamma$/hadron separation.

\section{Methods}

Cosmic rays, because of their electric charge, are deflected by ambient electromagnetic fields. As a result, it is very difficult to determine the origin of cosmic rays observed on Earth. This is not the case for $\gamma$-rays, which travel in straight lines from their sources, allowing us to study those sources indirectly. However, both cosmic rays (mainly protons) and $\gamma$-rays produce secondary particles that emit Cherenkov light, so both are detected by Cherenkov telescopes. The main challenge, which determines the sensitivity of the array, is our ability to effectively distinguish between signals from $\gamma$-rays and protons.

\begin{figure}[htbp]
  \centering
  \includegraphics[trim=0pt 0pt 0pt 0pt, clip, width=\linewidth]{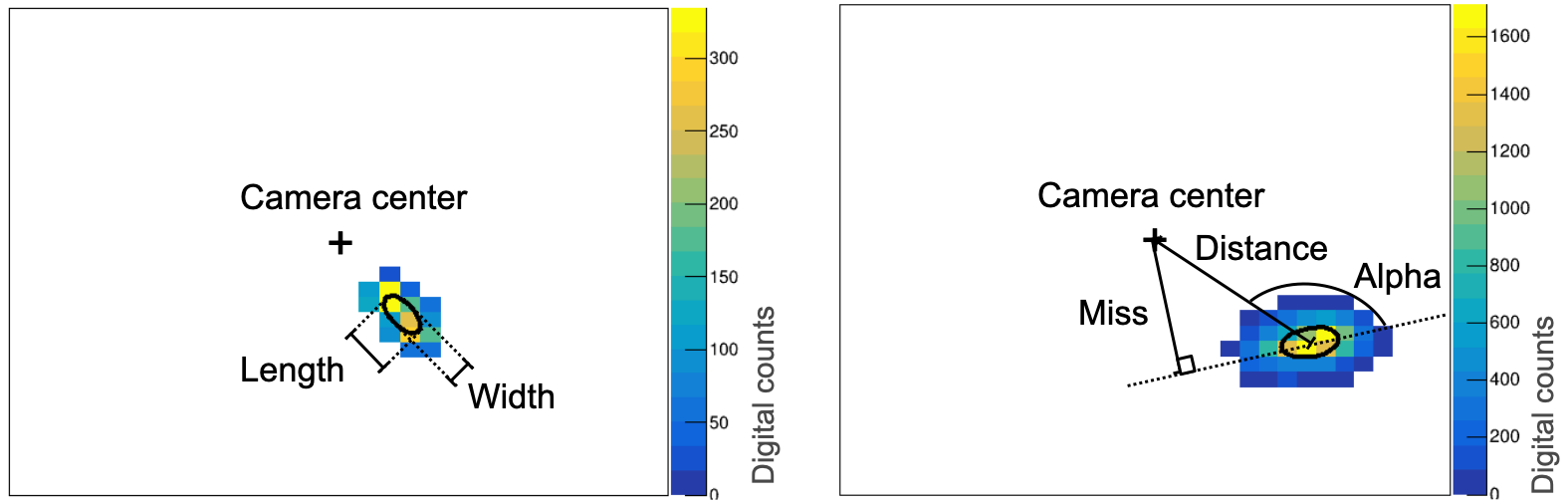}
  \caption{Example of a shower image formed on one telescope in units of digital counts of a \textbf{Left:} 100 TeV $\gamma$-ray and \textbf{Right:} a 200 TeV proton.}
  \label{fig:telimage}
\end{figure}

It is well known that the images recorded by the telescopes differ in several parameters, a group of which are known as the Hillas parameters \cite{hil85}. More complex methods for $\gamma$/hadron separation exist, but at this stage of the Dark100 array’s development, we focus on using Hillas parameters from single telescope images as a preliminary test of the effectiveness of $\gamma$/hadron separation. Fig.~\ref{fig:telimage} shows examples of a proton and a $\gamma$-ray shower image formed on a camera plane.

We use CORSIKA 7.8 to simulate proton and $\gamma$-ray showers, employing all four high-energy hadronic interaction models (QGSJET-III-01, DPMJET III, EPOS LHC-R, and SIBYLL 2.3e) and UrQMD 1.3\_cors for the low-energy hadronic interactions \cite{cor98}. Although all four models are expected to have a roughly similar behaviour, the differences in the specific description of the hadronic processes are one of the major sources of uncertainty in the hadron simulations. Following a similar study for CTA \cite{ohi21}, we also test the effect of the interaction model on $\gamma$/hadron separation. The $\gamma$-rays are simulated as coming from a point source at the center of the camera plane, while protons are modeled as a diffuse emission within a viewcone with an outer radius of 8$^\circ$, covering the diagonal of the camera plane. Both are made to originate at 20$^\circ$ zenith and 0$^\circ$ azimuth. For clarity, we use mono-energetic distributions at four different $\gamma$ energies: 10, 30, and 100 TeV, and 1 PeV. As the energy reconstruction is yet to be developed, here we assume for comparability, that the proton energy is twice the $\gamma$ energy. The CORSIKA showers are produced in a circular region of radii 400 m (for 10, 20, 30 and 60 TeV), 500 m (for 100 and 200 TeV) and 600 m (for 1 and 2 PeV) centred on the centre of gravity of the six-telescope array. The data are then analyzed for both a three-telescope and a six-telescope array. We present here the results for the three-telescope array.

The image parameters of interest, derived from the elliptical images formed on the camera plane are Hillas parameters, and a few related quantities:
\begin{itemize}
    \item Length and width: The length (semi-major axis) and width (semi-minor axis) of the ellipse. Ellipses formed by $\gamma$-ray showers are expected to be shorter and narrower than those formed by proton showers
    \item Miss: The shortest perpendicular distance from the camera center to the major axis of the ellipse. Since $\gamma$-ray showers are centered on the camera center (as we observe point sources), the $\gamma$-ray ellipses are oriented towards the camera center. As a result, the miss distance is shorter for $\gamma$-rays than for protons.
    \item Distance, maximum distance, and shower core distance: The distance between the camera center and the image center of gravity (in $^\circ$) is consistently small for $\gamma$-rays for the reason stated above. The maximum distance between a telescope and the event is always larger for protons than for $\gamma$-rays. The shower core distance refers to the distance between the camera plane and the impact location of the shower in meters.
    \item Alpha: This angle describes the orientation of the ellipse’s major axis with respect to the line connecting the camera center and the image center of gravity. Because $\gamma$-ray images are oriented towards the source position (which usually falls on the center of the camera), this angle is close to zero for $\gamma$-rays but can take a range of values for protons.
    \item Size: The total photon yield of the image in integrated digital counts across pixels.
\end{itemize}

Once these parameters are extracted from the shower images, machine learning algorithms can be used to consistently distinguish between protons and $\gamma$-rays. These algorithms are capable of combining various parameters into a single variable that indicates how proton-like or $\gamma$-like an event is. Following \cite{kra17}, we use \textit{Boosted Decision Trees} (BDTs). This algorithm is based on decision trees, which are a series of binary splits (e.g., Length > 0.5? Yes or No) that help reject background or proton signals. The decision tree is trained on a sample set of events where the type of event (0 for a proton and 1 for a $\gamma$) is known. \textit{Boosting} is a method that reduces error by reorganizing the tree iteratively.

Once the classifier is trained on a sample, it is then tested on a separate set of events. The accuracy of the trained BDT classifier can be evaluated in several ways. We consider two types of decision functions (one based on \texttt{predict\_proba} and one based on \texttt{decision\_function}). A decision function assigns a score to each event, and it should be possible to find a threshold score \textit{s} such that events with a score below \textit{s} are classified as protons, and those above as $\gamma$-rays. Ideally, these functions should show a clear separation between proton-like and $\gamma$-like events. 

Additionally, we also plot the Receiver Operating Characteristic (ROC) curve and the corresponding Area Under the Curve (AUC). The ROC curve plots the true positive rate (tpr) as a function of the false positive rate (fpr), illustrating the trade-off between correctly identifying $\gamma$-rays and misclassifying protons as $\gamma$-rays. If events were classified randomly, the ROC curve would be a straight line with a slope of 1. An ideal classifier would show a steep rise from (0, 0) to (0, 1) and then remain flat until (1, 1). As the classifier’s performance improves, the AUC approaches 1.

\section{Results}

\begin{figure}[htbp]
  \centering
  \includegraphics[width=\linewidth]{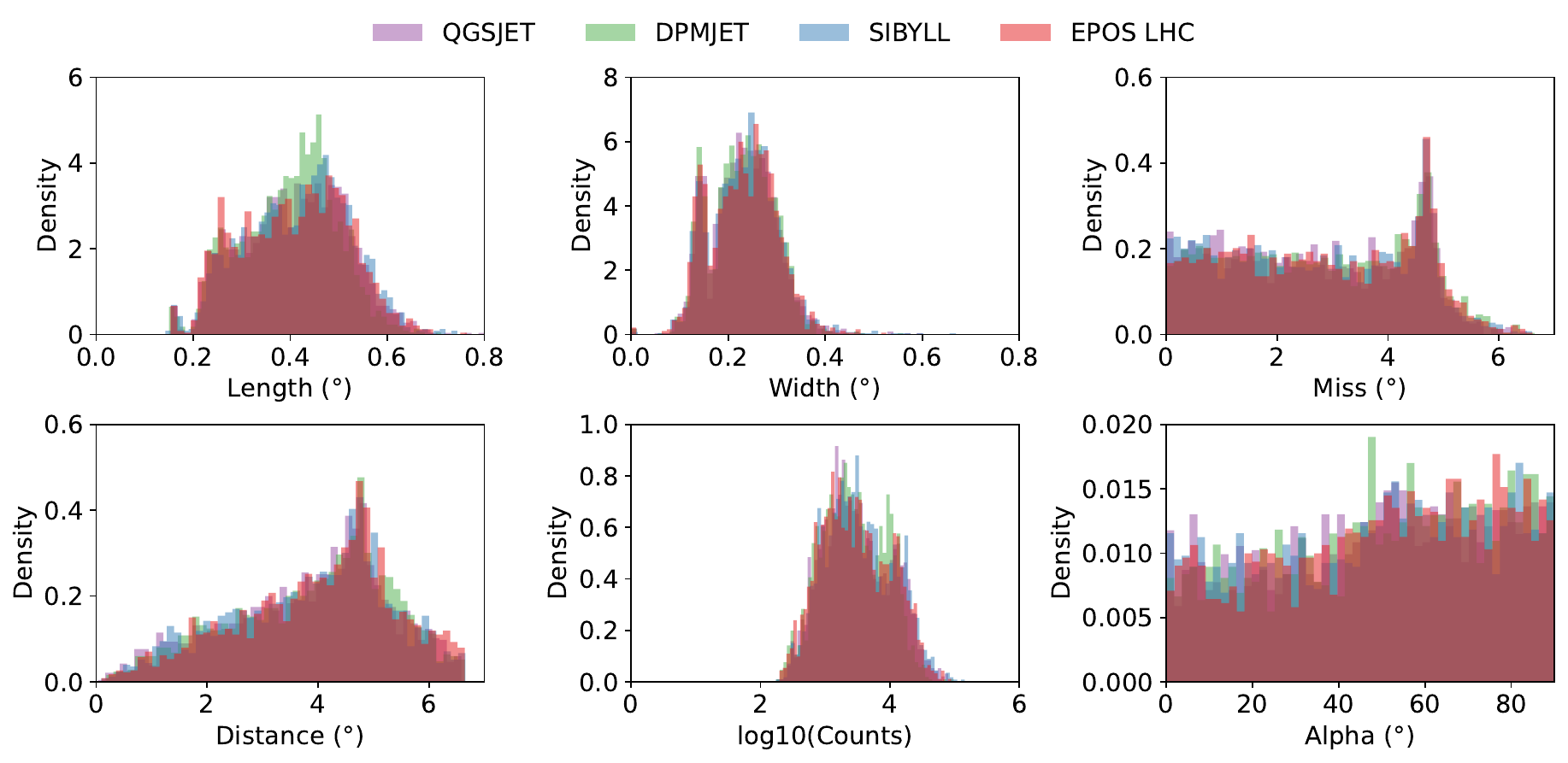}
  \caption{Image parameter distributions for 200 TeV proton showers generated using different hadronic interaction models.}
  \label{fig:hadmod}
\end{figure}

The effect of the hadronic interaction model was the first parameter tested. Fig.~\ref{fig:hadmod} shows the image parameter distributions for 200 TeV protons generated with different hadronic interaction models. Across the four energy bins, with 5000 diffuse proton showers generated in each case, there was no significant difference in the proton image parameters. Therefore, in the following sections, we present results using the QGSJET model.

\begin{figure}[htbp]
  \centering
  \includegraphics[width=\linewidth]{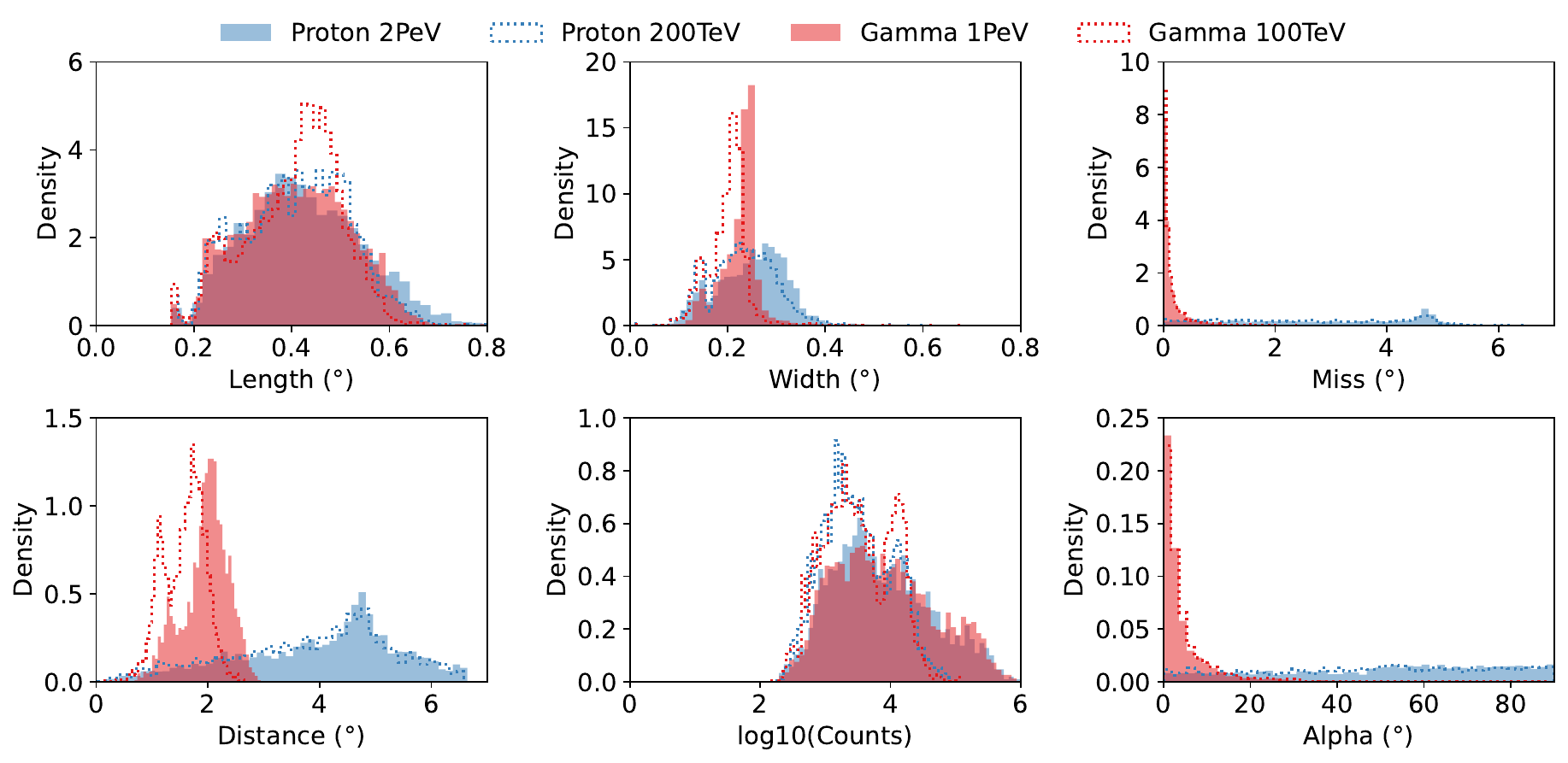}
  \caption{Image parameter distributions for proton (200 TeV and 2 PeV) and $\gamma$-ray (100 TeV and 1 PeV) showers produced using QGSJET hadronic interaction model.}
  \label{fig:engmod}
\end{figure}

The image parameters were then compared between showers from $\gamma$-rays with energy $E$ and protons with energy $2E$. Fig.~\ref{fig:engmod} shows the image parameter distributions for protons and $\gamma$-rays at $\gamma$ energies of 100 TeV and 1 PeV. The distributions of some parameters, such as miss, distance, and alpha, show a clear distinction between protons and $\gamma$-rays. The width parameter remains quite mixed at lower energies, but the peaks gradually shift at higher energies. The peaks in the length and size distributions also shift with increasing energy, but not enough to allow a clear separation between protons and $\gamma$-rays.

\begin{figure}[htbp]
  \centering
  \includegraphics[width=0.8\linewidth]{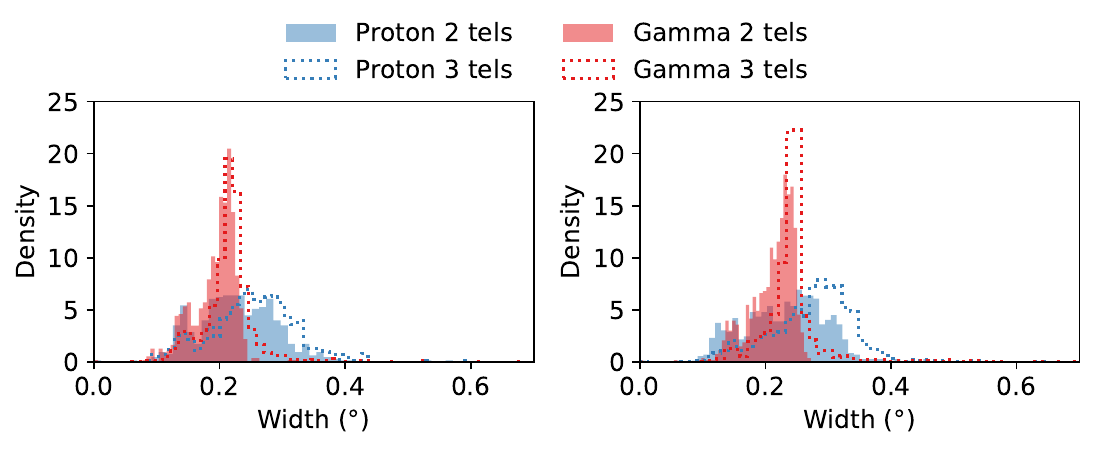}
  \caption{Image width distributions for proton and $\gamma$-ray at \textbf{Left:} 100 TeV and \textbf{Right:} 1 PeV, produced using QGSJET hadronic interaction model for events triggering 2 and 3 telescopes.}
  \label{fig:telmod}
\end{figure}

The peaks shift not only with energy but also with the number of triggered telescopes. Fig.~\ref{fig:telmod} shows the width distributions for proton and $\gamma$-ray showers for $\gamma$-ray energies 100 TeV (on the left panel) and 1 PeV (on the right panel), where the number of telescopes triggered by the event can be 2 or 3. As the number of triggered telescopes increases, there may be a better chance of separating protons from $\gamma$-rays.

\begin{figure}[htbp]
  \centering
  \includegraphics[width=0.69\linewidth]{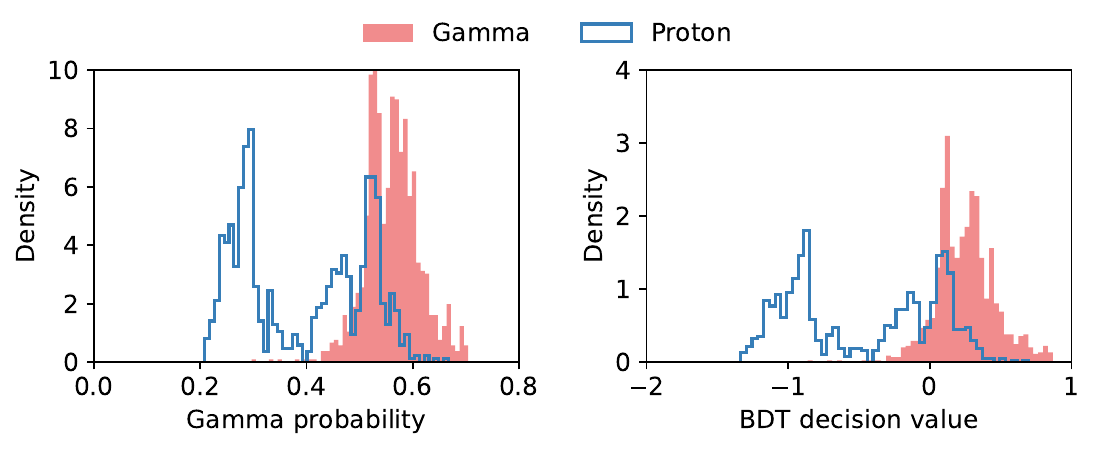}
  \includegraphics[width=0.38\linewidth]{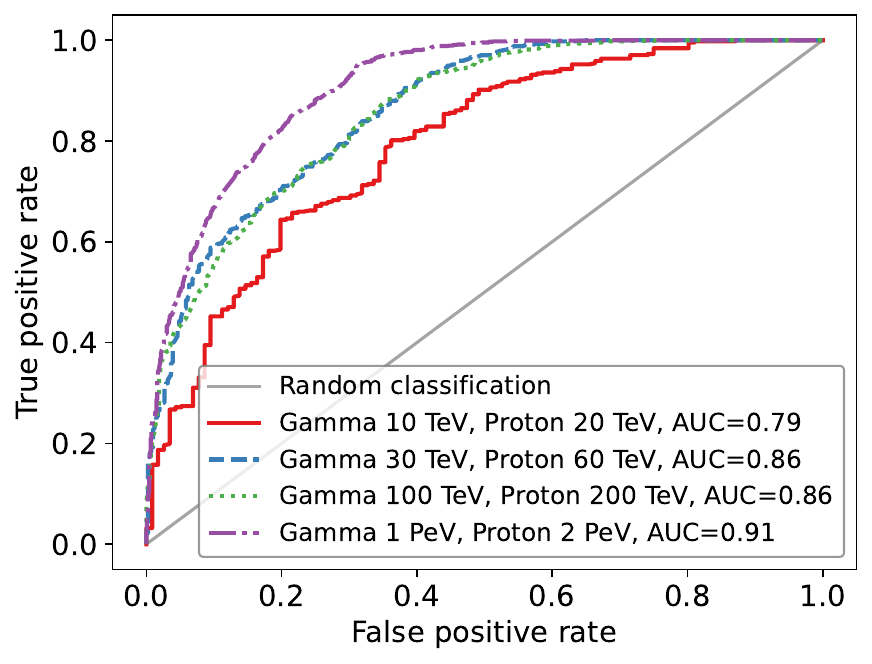}
  \caption{Performance tests on the BDTs for the case of 1 PeV $\gamma$-rays (and 2 PeV protons) produced using QGSJET.}
  \label{fig:bdt}
\end{figure}

Finally, the BDTs trained on a sample of this data using length, width and size were tested on the remaining data and the results for 1 PeV $\gamma$-rays and 2 PeV protons are shown. The two top panels of Fig.~\ref{fig:bdt} show the probability of an event being $\gamma$-like, and the decision function. Despite some overlap between protons and $\gamma$-rays, there is an approximate boundary between the score for protons and those for $\gamma$-rays. The bottom panel of Fig.~\ref{fig:bdt} shows the ROC curve for the case of 1 PeV $\gamma$-ray showers. The corresponding AUC is 0.91, which is reasonably close to 1.

\section{Discussion \& Conclusion}

Dark100 is an array of PANOSETI telescopes which will be located at the Palomar Observatory. This array of Fresnel lens telescopes will be sensitive to Cherenkov light from ultra-high-energy $\gamma$-rays and cosmic rays between 100 TeV and above 1 PeV. In this study, we have carried out a preliminary test on the quality of $\gamma$/hadron separation using the Hillas parameters from single-telescope images. After parameterization, we train BDTs on a sample of events, focusing on a subset of parameters (size, length, and width). The performance of the BDTs is evaluated using decision function curves and ROC curves.

We find that the choice of high-energy hadronic interaction model does not significantly affect the image parameter distributions for protons. Some image parameters provide a clear distinction between protons and gammas, while others are less effective. The separation improves as the energy and the number of triggered telescopes increase.

The results from both the \texttt{predict\_proba} and \texttt{decision\_function} outputs of the BDT show that, although there is some overlap between proton and gamma scores, the two can be effectively separated.  The ROC curves further demonstrate that this separation is significantly better than random classification, with AUC values above 0.9 at the highest energies.

However, this separation is expected to improve as this is only a preliminary study. At the moment, the telescope is still undergoing lab tests and a complete characterization of the telescope response is pending. Therefore, the telescope model, the point spread function and the noise calculations used in this study are all approximations. The energy reconstruction is also in development. Moreover, all $\gamma$/hadron separation is currently based on single telescope image parameters. The separation is expected to improve with more complex $\gamma$/hadron separation algorithms. In the future, the study is set to be repeated with more statistics, the use of mean-scaled reduced parameters, more precise knowledge of the telescope and a more advanced method of $\gamma$/hadron separation.

\acknowledgments
\small{The PANOSETI program is a collaborative initiative involving researchers and instrumentation teams from UC San Diego, UC Berkeley, Harvard University, Caltech, Ruhr University Bochum, and the University of Delaware. It is made possible through the generous support of Franklin Antonio and the Bloomfield Family Foundation. S. Ravikularaman is supported by an ERC Consolidator Grant (No. 101124914).}

\bibliography{refs}

\clearpage

\section*{PANOSETI Collaboration}

\begin{description}
\item[Aaron Brown]{Department of Astronomy \& Astrophysics, University of California San Diego, USA}
\item[Benjamin Godfrey]{Department of Astronomy, University of California Berkeley, USA}
\item[Jamie Holder]{Department of Physics and Astronomy and the Bartol Research Institute, University of Delaware, USA}
\item[Nikolas Korzoun]{Department of Physics and Astronomy and the Bartol Research Institute, University of Delaware, USA}
\item[Andrew Howard]{Astronomy Department, California Institute of Technology, USA}
\item[Paul Horowitz]{Department of Physics, Harvard University, USA}
\item[Alyssa Johnson]{Department of Astronomy \& Astrophysics, University of California San Diego, USA}
\item[Wei Liu]{Department of Astronomy, University of California Berkeley, USA}
\item[Jerome Maire]{Department of Astronomy \& Astrophysics, University of California San Diego, USA}
\item[Yuriy Popovych]{Astronomisches Institut (AIRUB), Ruhr-Universität Bochum, Germany}
\item[Elisa Pueschel]{Astronomisches Institut (AIRUB), Ruhr-Universität Bochum, Germany}
\item[Nicolas Rault-Wang]{Department of Astronomy, University of California Berkeley, USA} 
\item[Sruthiranjani Ravikularaman]{Astronomisches Institut (AIRUB), Ruhr-Universität Bochum, Germany}
\item[Dan Werthimer]{Department of Astronomy, University of California Berkeley, USA}
\item[Shelley Wright]{Department of Astronomy \& Astrophysics, University of California San Diego, USA}
\end{description}

\end{document}